\documentclass[journal]{IEEEtran}
\IEEEoverridecommandlockouts
\usepackage{cite}
\usepackage{amsmath, amssymb, amsfonts, mathtools, bm, amsthm}
\usepackage{blkarray}
\allowdisplaybreaks[4]
\usepackage{algorithmic}
\usepackage{graphicx}
\usepackage{textcomp}
\usepackage{xcolor}
\usepackage{placeins}
\usepackage{makecell}
\usepackage{standalone}
\allowdisplaybreaks
\usepackage{soul}
\usepackage{stfloats}
\usepackage{tabularx}
\usepackage{booktabs}
\usepackage{array}
\usepackage{hyperref}
\usepackage{url}
\usepackage{graphicx} % Required for \rotatebox
\usepackage{diagbox}
\usepackage[ruled,vlined,linesnumbered]{algorithm2e} % for algorithm environment
\usepackage{xcolor} % for colored titles
\usepackage[ruled,vlined]{algorithm2e}
\def\BibTeX{{\rm B\kern-.05em{\sc i\kern-.025em b}\kern-.08em
 T\kern-.1667em\lower.7ex\hbox{E}\kern-.125emX}}
\usepackage{standalone}
\usepackage{tikz}\usetikzlibrary{calc, shapes, arrows, automata, positioning, patterns}

\usepackage{hyperref}
\hypersetup{
    colorlinks=true,      % True: colored links; False: boxed links
    linkcolor=blue,       % Color of internal links (sections, pages, etc.)
    citecolor=green,      % Color of citation links
    filecolor=magenta,    % Color of file links
    urlcolor=cyan         % Color of external URL links
}

\begin{document}
% \title{Beyond Yes or No: Grading Observability of Power System Dynamic States with Graph-Based Symbolic Coefficients}

% \title{Measurement Selection Using Graph-based Observability for Power System Dynamic State Estimation}
\title{Beyond Yes or No: Grading Observability of Power System Dynamic States with Symbolic Coefficients}
\author{Akhila~Kandivalasa,~\IEEEmembership{Student~Member,~IEEE,}~Marcos~Netto,~\IEEEmembership{Senior~Member,~IEEE}
\thanks{This work is supported by the National Science Foundation under Grant 2328241. The authors are with the Department of Electrical and Computer Engineering, New Jersey Institute of Technology, Newark, NJ 07102, USA.}
}

\maketitle

\begin{abstract}
Graph-based observability analysis runs in seconds where Lie-derivative tests take hours, but it answers only one question: is the system observable or not. It cannot say how well a given set of measurements supports state reconstruction. This paper extends the graph-based approach with symbolic observability coefficients computed from a symbolic Jacobian of the state-space equations, in which every dependency is classified as constant, nonlinear polynomial, or rational. We show that the coefficient tracks the analytical observability determinant of the Lie-derivative approach: it is zero when the determinant vanishes, lies strictly between zero and one when the determinant depends on the state, and equals one when the determinant is a nonzero constant. We apply the method to a seven-state synchronous machine and to the 21-state Western System Coordinating Council nine-bus system under several synchrophasor measurement configurations. On the nine-bus system, the coefficient grows from 0.0952 with one PMU to 0.1905 with two and 0.2857 with three, and the whole analysis runs in under five seconds. Configurations with the same number of PMUs receive the same coefficient, yet an unscented Kalman filter estimates the states with different accuracy for each. The symbolic coefficient therefore grades measurement sets beyond identifying root strongly connected components, while telling PMU locations apart still requires the numerical parameters of the system.
\end{abstract}


\section{Introduction}

Dynamic state estimation (DSE) \cite{Zhao2026} requires observability. For a linear system, this is settled by a rank test \cite{KALMAN1960491}. For the nonlinear models that describe power system dynamics, it is still an open problem \cite{Zhao2019}. The classical extension of the rank test to nonlinear systems relies on Lie derivatives \cite{Hermann_Krener_1977} and does not scale. On the Western System Coordinating Council (WSCC) nine-bus system, a model with only 21 state variables \cite[pg. 171]{Sauer1997}, the Lie-derivative test takes about two hours \cite{Kandivalasa2026}.

Graph-based observability analysis \cite{Liu2013} sidesteps that cost. State variables become nodes, and dependencies between them become directed edges. The root strongly connected components (SCCs) of the resulting digraph then mark the observability condition: measuring at least one variable in each root SCC. The test reads only which variable appears in which equation and never touches a parameter value. Computing root SCCs scales linearly in the number of nodes and edges, so the analysis runs in seconds \cite{Kandivalasa2026} for the same WSCC test case. The answer it gives, however, is binary: observable or not.  

That answer is incomplete for nonlinear systems, where observability varies across the state space. It depends on whether the coupling between state variables is linear or nonlinear, and on the node dynamics themselves \cite{letellier2009symbolic}. One wants to know how strongly a system is observable and from which measurements. Rouhani and Abur \cite{Rouhani2017} rank the state variables of a two-axis synchronous machine model by the singular values of the Lie-derivative observability matrix, and pay the full cost of the Lie derivatives. Qi, Sun, and Kang \cite{Qi2015} rank phasor measurement unit (PMU) locations with the empirical observability Gramian instead. It needs no derivatives, but it is built from repeated nonlinear simulations and so depends on parameter values and the operating point. The graph-based approach needs a way to quantify observability that keeps its speed. 

Symbolic observability coefficients offer one. Letellier and Aguirre introduced them for low-dimensional chaotic systems \cite{letellier2009symbolic}, and Letellier and co-workers later extended them to networks \cite{letellier2018symbolic}. The idea is to replace each entry of the Jacobian with a symbol for its type: constant, nonlinear polynomial, or rational. The observability matrix is then built from those symbols rather than from the expressions themselves, and its determinant collapses into a single coefficient $\eta$ between zero and one. The higher $\eta$, the fewer the singularities in the state space where observability is lost. To our knowledge, these coefficients have not been applied to power system models before.

This paper brings symbolic observability coefficients into the graph-based approach to grade observability within a root SCC. On a seven-state synchronous machine model, the coefficient ranks the four root SCC variables and tracks the analytical determinant of the Lie-derivative approach, equal to one when that determinant is a nonzero constant. We then carry the grading over to the terminal voltage phasors that PMUs provide on the WSCC nine-bus system and test the selected configurations with an unscented Kalman filter (UKF). With dependencies entered as symbols rather than expressions, the full analysis of the 21-state WSCC system runs in under five seconds.

This paper proceeds as follows. Section \ref{sec:observability} reviews nonlinear observability and the graph-based test. Section \ref{sec:symbolic} introduces symbolic observability coefficients and applies them to measurement selection. Section \ref{sec:results} reports results on the WSCC nine-bus system. Section \ref{sec:conclusion} concludes the paper and outlines directions for future work.

\section{Observability of Nonlinear Systems}
\label{sec:observability}
Consider a nonlinear dynamical system,
\begin{equation}\label{eq:nonlinearsystem}
\bm{\dot{x}}(t)=\bm{f}(\bm{x}(t),\bm{u}(t)), \quad \bm{y}(t)=\bm{h}(\bm{x}(t)),
\end{equation}
evolving on a smooth manifold $\mathcal{M}\subseteq\mathbb{R}^{n}$, where $\bm{x}\in\mathcal{M}$, $\bm{u}\in\mathbb{R}^{m}$, and $\bm{y}\in\mathbb{R}^{r}$ denote the state, input, and measured output vectors, respectively. $\bm{f}$ is the vector field governing the system dynamics, while $\bm{h}$ is the measurement function mapping the system states to the measured outputs.

The system \eqref{eq:nonlinearsystem} is observable if its initial state $\bm{x}_0$ can be uniquely determined from its measured output $\bm{y}$, given known inputs $\bm{u}$. The classical test computes the rank of an observability matrix built from Lie derivatives of the output \cite{Hermann_Krener_1977}. The Lie derivative ${\mathcal{L}}_{\bm{f}} \bm{y}$ measures how the output changes as the state moves along the vector field $\bm{f}$,
\begin{equation}\label{lie}
\mathcal{L}_{\bm{f}}\bm{y} \coloneqq \frac{\partial \bm{h}(\bm{x})}{\partial x_1} \dot{x}_1 + \frac{\partial \bm{h}(\bm{x})}{\partial x_2} \dot{x}_2 + \dots + \frac{\partial \bm{h}(\bm{x})}{\partial x_n} \dot{x}_n 
\end{equation}
and higher orders follow recursively,
\begin{equation}\label{eq:lie}
\mathcal{L}_{\bm f}^{k}\bm{y}=\frac{\partial(\mathcal{L}_{\bm f}^{k-1}\bm{y})}{\partial\bm{x}}\bm{f},\quad k\geq1, \quad \mathcal{L}_{\bm f}^{0}\bm{y}=\bm{h}.
\end{equation}
Each successive Lie derivative reveals dependencies on state variables that are not explicitly present in the original output functions. Thus, the output functions and their Lie derivatives may provide sufficient information to uniquely reconstruct the system state. Collecting them gives the map
\begin{equation}\label{eq:outputmap}
\bm{g}(\bm{x})=\begin{bmatrix}\bm{h}^{\top} & (\mathcal{L}_{\bm f}\bm{y})^{\top} & \cdots & (\mathcal{L}_{\bm f}^{n-1}\bm{y})^{\top}\end{bmatrix}^{\top}
\end{equation}
whose Jacobian with respect to $\bm{x}$ is the observability matrix,
\begin{equation}
\mathcal{O} \coloneqq \frac{\partial{\bm{g}(\bm{x})}}{\partial\bm{x}}\bigg|_{\bm{x}=\bm{x}_{0}}
\end{equation}
If $\mathcal{O}$ has full column rank $n$, the outputs and their derivatives carry enough independent information to reconstruct $\bm{x}$, and the system \eqref{eq:nonlinearsystem} is observable.\footnote{Strictly, following \cite{Hermann_Krener_1977}, if rank$\mathcal{O}=n$, \eqref{eq:nonlinearsystem} is said \emph{locally weakly observable} at $\bm{x}_0$. In this paper, we simply say ``observable."}

The rank test gives a binary answer, i.e., whether a system is observable or not. Rouhani and Abur \cite{Rouhani2017} leverage the singular value decomposition of $\mathcal{O}$ to obtain a measure of \emph{how observable} \eqref{eq:nonlinearsystem} is. The larger the smallest singular value of $\mathcal{O}$ is, the stronger the observability. Both the rank test and the singular value decomposition of $\mathcal{O}$ rely on heavy symbolic computation and do not scale to power grids of any relevant size. The graph-based approach in \cite{Kandivalasa2026} replaces that computation with a directed graph (digraph) built from the dependency structure of \eqref{eq:nonlinearsystem}.

Let $\mathcal{D}=(\mathcal{V},\mathcal{E})$ be a digraph constructed from the system Jacobian $\bm{J}=\partial\bm{f}/\partial\bm{x}$. Each state variable yields a node, and a directed edge from $x_i$ to $x_j$, denoted $x_i\rightarrow x_j$, exists whenever the equation for $\dot{x}_i$ in \eqref{eq:nonlinearsystem} explicitly contains the state $x_j$. The corresponding entries of a binary adjacency matrix are
\begin{equation}\label{eq:adjacency}
a_{ij}=
\begin{cases}
1, & \bm{J}_{ij}\not\equiv 0,\\
0, & \bm{J}_{ij} \equiv 0.
\end{cases}
\end{equation}

Let us consider an example. Take the nonlinear system
\begin{equation}
\dot{x}_1 = x_2^2 - x_1 + x_4,\; 
\dot{x}_2 = x_3,\;
\dot{x}_3 = x_1^2,\;
\dot{x}_4 = x_4. \label{eq:nonlinear_exampl}
\end{equation}
Fig.\ref{fig:digraphexample} shows its digraph and adjacency matrix.
\begin{figure}[!t]
\centering
\begin{minipage}[c]{0.48\columnwidth}
    \centering
    \begin{tikzpicture}[node distance={13mm}, thick, main/.style = {draw, circle}] 
% \filldraw[densely dotted, color=red!60, fill=red!15, very thick, opacity=0.4] (-.5,-0.45) rectangle (5,.8);
% \filldraw[densely dotted, color=blue!60, fill=blue!15, very thick, opacity=0.4] (-3,-0.45) rectangle (-.8,.8);
%
% \draw (3.5,.8) node[anchor=north west] {\small root SCC};
% \draw (-3,.8) node[anchor=north west] {\small SCC};
%
\node[main] (1) [            fill=white] {$x_{1}$};
\node[main] (2) [ left of=1, fill=white] {$x_{4}$};
\node[main] (3) [right of=1, fill=white] {$x_{2}$};
\node[main] (4) [right of=3, fill=white] {$x_{3}$};
% \node[main] (5) [right of=4, color=white] {}; % spacer
%
\draw[->] (1) to [out=135,in=85,looseness=3.5] (1);
\draw[->] (2) to [out=135,in=85,looseness=3.5] (2);
\draw[->] (4) to [out=135,in=45,looseness=0.8] (1);
\draw[->] [color=black] (1) -- (2);
\draw[->] [color=black] (1) -- (3);
\draw[->] [color=black] (3) -- (4);
\end{tikzpicture}
\end{minipage}
\hfill
\begin{minipage}[c]{0.49\columnwidth}
    \centering
    \begingroup
    \footnotesize
    \setlength{\arraycolsep}{2pt}
    \renewcommand{\arraystretch}{0.9}
    $\mathbf{A} =
    \begin{blockarray}{ccccc}
    {\color{gray}x_1} & {\color{gray}x_2} & {\color{gray}x_3} & {\color{gray}x_4} & \\
    \begin{block}{(cccc)c}
        1 & 1 & 0 & 1 & {\color{gray}x_1} \\
        0 & 0 & 1 & 0 & {\color{gray}x_2} \\
        1 & 0 & 0 & 0 & {\color{gray}x_3} \\
        0 & 0 & 0 & 1 & {\color{gray}x_4} \\
    \end{block}
    \end{blockarray}$
    \endgroup
\end{minipage}
\vspace{-.3cm}
\caption{Digraph and corresponding adjacency matrix for \eqref{eq:nonlinear_exampl}.}
\label{fig:digraphexample}
\end{figure}

The digraph is partitioned into SCCs, maximal subgraphs in which every node can be reached from every other node. An SCC with no incoming edges is a root SCC. Fig. \ref{fig:digraphexample} has two SCCs, $\{x_1,x_2,x_3\}$ and $\{x_4\}$; the first is a root SCC. A system is observable if at least one variable from each root SCC is measured, because that measurement reaches the state dependencies of all remaining components. Successive Lie derivatives expose the same dependencies, and \cite{Kandivalasa2026} shows that the two approaches agree. Root SCCs are computed with standard graph algorithms on the adjacency matrix in a few seconds. In fact, cutting-edge numerical methods for computing root SCCs run in linear time with respect to the number of nodes and edges in a graph, so the graph approach scales to power grids of significant size. This contrasts sharply with the approach based on Lie derivatives.

Two gaps remain. The graph approach in \cite{Kandivalasa2026} identifies the root SCC but says nothing about which variable or variables inside it should be measured, and it offers no metric comparable to the singular value decomposition of $\mathcal{O}$. The symbolic observability coefficient of Letellier and Aguirre \cite{letellier2009symbolic} closes the first gap and, in doing so, the second. The next section leverages the symbolic observability coefficient to identify the interplay between PMU selection and observability strength.

\section{Symbolic Observability Coefficients for Measurement Selection}
\label{sec:symbolic}
The Lie derivatives \eqref{lie}-\eqref{eq:lie} of Section \ref{sec:observability} describe \eqref{eq:nonlinearsystem} through its measured outputs and their successive Lie derivatives along the vector field $\bm{f}$. The state variables span the original state space; the derivatives span a reconstructed space. The map between the two is a differential embedding \cite{Letellier2005Relation}.

For $\bm{y}\in\mathbb{R}^r$, the reconstructed space $\bm{\Phi}_{\bm{y}}:\mathcal{M}\rightarrow\mathbb{R}^{d_r}$ is
\begin{equation}\label{eq:embedding}
\bm{\Phi}_{\bm{y}}=
\begin{bmatrix}y_1 & \dots & \mathcal{L}_{\bm f}^{k_1}y_1 & \dots & y_r & \dots & \mathcal{L}_{\bm f}^{k_r}y_r\end{bmatrix}^{\top},
\end{equation}
where $d_r=\sum_{i=1}^{r}(k_i+1)\geq n$ is its dimension. By the inverse function theorem, the map is locally invertible when its Jacobian, the observability matrix, has full column rank $n$. Full rank at one point is only part of the picture. Which outputs are measured, and how many derivatives of each enter $\bm{\Phi}_{\bm{y}}$, decide where in the state space that rank holds.

Let us return to the example system \eqref{eq:nonlinear_exampl}. With a single output $y$, three successive Lie derivatives are needed to reach $d_r=4$. Suppose $y=x_4$. Every Lie derivative of $y$ along $\bm{f}$ depends only on $x_4$, so the observability matrix has rank one and $\det\mathcal{O}_{(x_4,\dot{x}_4,\ddot{x}_4,x_4^{(3)})} = 0$. The system is unobservable. Now, let $y=x_2$. Its first three derivatives give a full-rank matrix with $\det\mathcal{O}_{(x_2,\dot{x}_2,\ddot{x}_2,x_2^{(3)})} = 4x_1^2$. The determinant depends on the state and vanishes at $x_1=0$, so the system is observable everywhere except on that singular set. Finally, suppose we measure both $x_1$ and $x_2$. Only two derivative coordinates are needed for $d_r=4$, and $\det\mathcal{O}_{(x_1,x_2,\dot{x}_1,\dot{x}_2)}=-1$. Observability holds throughout the state space.

All three cases reach $d_r=4$. What separates them is the choice of measured variables and the derivatives taken of each. The ideal configuration gives $\mathcal{O}$ full column rank $n$ over the whole admissible state space, with a constant determinant. Symbolic observability coefficients \cite{letellier2018symbolic} estimate how close a configuration comes to that ideal without computing the determinant analytically.

\subsection{Symbolic Observability Coefficients}
To obtain a symbolic coefficient, we start from a symbolic version of the system Jacobian ($\bm{J}=\partial\bm{f}/\partial\bm{x}$). Each entry of $\bm{J}$ is classified as zero ($0$), constant ($1$), nonlinear polynomial ($\bar{1}$), or rational ($\bar{\bar{1}}$). The adjacency matrix $\bm{A}$ in Fig.\ref{fig:digraphexample} records only whether an entry is zero, whereas the symbolic Jacobian retains its type. For the example system \eqref{eq:nonlinear_exampl}, the symbolic Jacobian is
\begin{equation}\label{eq:symbolicjac}
\widetilde{\bm{J}}=
\begin{bmatrix}
1 & \bar{1} & 0 & 1\\
0 & 0 & 1 & 0\\
\bar{1} & 0 & 0 & 0\\
0 & 0 & 0 & 1
\end{bmatrix}.
\end{equation}

For a measured output \(y=x_i\), the symbolic observability matrix \(\widetilde{\mathcal O}\) is initialized with the first row \(\widetilde{\mathcal O}_{1,:}=\bm{e}_i^\top\), representing the direct measurement, and the second row \(\widetilde{\mathcal O}_{2,:}=\widetilde{\bm{J}}_{i,:}\), representing its first Lie derivative. Each subsequent row is computed recursively as follows:
\begin{equation}\label{eq:symbolicrecursion}
\widetilde{\mathcal{O}}_{k,j}=
\bigoplus_{\ell=1}^{n}
\left(\widetilde{\mathcal{O}}_{k-1,\ell}
\otimes\widetilde{\bm{J}}_{\ell,j}\right),
\quad k\geq3,
\end{equation}
where \(k\) denotes the row index (corresponding to the \((k-1)\)th derivative), \(j\) the state-variable index, \(\ell\) is the summation index over the \(n\) states, and $\otimes$ and $\oplus$ denote, respectively, symbolic multiplication and addition based on the following rules:
\begin{equation}\label{eq:symbolicrules}
\begin{gathered}
\begin{array}{c|cccc}
\otimes & 0 & 1 & \bar{1} & \bar{\bar{1}}\\ \hline
0 & 0 & 0 & 0 & 0\\
1 & 0 & 1 & \bar{1} & \bar{\bar{1}}\\
\bar{1} & 0 & \bar{1} & \bar{1} & \bar{\bar{1}}\\
\bar{\bar{1}} & 0 & \bar{\bar{1}} & \bar{\bar{1}} & \bar{\bar{1}}
\end{array}
\qquad
\begin{array}{c|cccc}
\oplus & 0 & 1 & \bar{1} & \bar{\bar{1}}\\ \hline
0 & 0 & 1 & \bar{1} & \bar{\bar{1}}\\
1 & 1 & 1 & \bar{1} & \bar{\bar{1}}\\
\bar{1} & \bar{1} & \bar{1} & \bar{1} & \bar{\bar{1}}\\
\bar{\bar{1}} & \bar{\bar{1}} & \bar{\bar{1}} & \bar{\bar{1}} & \bar{\bar{1}}
\end{array}
\end{gathered}
\end{equation}

These operations retain the highest algebraic complexity among the contributing dependencies. For instance, measuring $x_2$ in \eqref{eq:nonlinear_exampl} gives $\widetilde{\mathcal{O}}_{1,:}=[0,1,0,0]$ and $\widetilde{\mathcal{O}}_{2,:}=[0,0,1,0]$. The recursion \eqref{eq:symbolicrecursion} then produces $\widetilde{\mathcal{O}}_{3,:}=[\bar{1},0,0,0]$ and $\widetilde{\mathcal{O}}_{4,:}=[\bar{1},\bar{1},0,\bar{1}]$, yielding
\begin{equation}\label{eq:symbolicobsx2}
\widetilde{\mathcal{O}}_{x_2}=
\begin{bmatrix}
0 & 1 & 0 & 0\\
0 & 0 & 1 & 0\\
\bar{1} & 0 & 0 & 0\\
\bar{1} & \bar{1} & 0 & \bar{1}
\end{bmatrix}.
\end{equation}

The determinant of the resulting matrix is evaluated using the same symbolic arithmetic, retaining the number of constant ($N_1$), polynomial ($N_{\bar{1}}$), and rational ($N_{\bar{\bar{1}}}$) symbols in its nonzero terms. Finally, the symbolic observability coefficient
\begin{equation}\label{eq:symboliccoeff}
\eta \coloneqq \frac{N_1}{D}+\frac{N_{\bar{1}}}{D^2}+\frac{N_{\bar{\bar{1}}}}{D^3},
\end{equation}
where $D=\max(1,N_1)+N_{\bar{1}}+N_{\bar{\bar{1}}}$ and $0\leq\eta\leq1$. Constant dependencies weigh the most. Polynomial and rational terms weigh progressively less because they can create singularities. A value of $\eta=1$ represents the ideal condition, i.e., a nonzero determinant that does not depend on the state and guarantees full rank everywhere. Smaller values signal growing nonlinear contributions and possible singular sets, and $\eta=0$ marks an unobservable configuration. In the example system \eqref{eq:nonlinear_exampl}, for $\widetilde{\mathcal{O}}_{x_2}$, the nonzero determinant term holds two constant and two polynomial symbols, so $N_1=2$, $N_{\bar{1}}=2$, $N_{\bar{\bar{1}}}=0$, and $\eta_{x_2}=\frac{2}{4}+\frac{2}{4^2}=0.625$. Similarly, the other two configurations give $\eta_{x_4}=0$ and $\eta_{(x_1,x_2)}=1$. These values track the analytical determinants: $\eta=0$ for a vanishing determinant, $0<\eta<1$ for a state-dependent nonzero determinant, and $\eta=1$ for a constant nonzero one. The ideal $\eta=1$ may not be attainable with the available measurements. Maximizing $\eta$ remains a useful criterion, since it favors state reconstructions with less nonlinear complexity and fewer potential singularities. Accordingly, we rank the measurement configurations that satisfy the root-SCC requirement of \cite{Kandivalasa2026} by their symbolic observability coefficient.
\vspace{-8pt}
\subsection{Selection of Root-SCC Variables}
\label{sec:rootsccselection}
Consider the two-axis model with an IEEE Type-1 exciter from \cite{Sauer1997}, with $R_s=0$ and terminal voltage $V_{Di}+jV_{Qi}$ treated as an input. All currents are expressed in the network reference frame; the stator algebraic equations give them explicitly as
\begin{subequations}\label{eq:machine}
\begin{align}
I_{Di} &= \tfrac{\sin\delta_i}{X'_{di}}\big(E'_{qi}-V_{Di}\cos\delta_i-V_{Qi}\sin\delta_i\big)
  \notag\\
 &\quad +\tfrac{\cos\delta_i}{X'_{qi}}\big(V_{Di}\sin\delta_i-V_{Qi}\cos\delta_i-E'_{di}\big),
  \notag\\
I_{Qi} &= -\tfrac{\cos\delta_i}{X'_{di}}\big(E'_{qi}-V_{Di}\cos\delta_i-V_{Qi}\sin\delta_i\big)
  \notag\\
 &\quad +\tfrac{\sin\delta_i}{X'_{qi}}\big(V_{Di}\sin\delta_i-V_{Qi}\cos\delta_i-E'_{di}\big),
\end{align}
and the state equations are
\begin{align}
T'_{d0i}\,\dot E'_{qi} &= -E'_{qi}+E_{fdi} \notag\\
 &\quad -(X_{di}-X'_{di})\big(I_{Di}\sin\delta_i-I_{Qi}\cos\delta_i\big),
  \notag\\
T'_{q0i}\,\dot E'_{di} &= -E'_{di} +(X_{qi}-X'_{qi})\big(I_{Di}\cos\delta_i+I_{Qi}\sin\delta_i\big),
  \notag\\
\dot\delta_i &= \omega_i-\omega_s,
  \notag\\
\tfrac{2H_i}{\omega_s}\,\dot\omega_i &= T_{Mi}-D_i(\omega_i-\omega_s)
  -\big(V_{Di}I_{Di}+V_{Qi}I_{Qi}\big),
  \notag\\
T_{Ei}\,\dot E_{fdi} &= -\big(K_{Ei}+S_{Ei}(E_{fdi})\big)E_{fdi}+V_{Ri},
  \notag\\
T_{Fi}\,\dot R_{fi} &= -R_{fi}+\tfrac{K_{Fi}}{T_{Fi}}\,E_{fdi},
  \notag\\
T_{Ai}\,\dot V_{Ri} &= -V_{Ri}+K_{Ai}R_{fi}-\tfrac{K_{Ai}K_{Fi}}{T_{Fi}}\,E_{fdi}
  \notag\\
 &\quad +K_{Ai}\big(V_{\mathrm{ref}i}-V_{ti}\big),
  \label{eq:dif}
\end{align}
\end{subequations}
where $V_{ti}=\sqrt{V_{Di}^2+V_{Qi}^2}$.
For a decentralized DSE setting, with $\bm{u} = [\, T_{Mi}\; V_{\textnormal{ref}i}\; V_{Di}\; V_{Qi} \,]^\top$ and $\bm{y} = [\, {I_{Di}}\; I_{Qi}\,]^\top,$ the model \eqref{eq:dif} takes the form $\bm{\dot{x}}= \bm{f}(\bm{x,u})$ of \eqref{eq:nonlinearsystem} with state vector $\bm{x}=[\,E'_q\;E'_d\;\delta\;\omega\;E_{fd}\;R_f\;V_R\,]^\top$. 
We treat the saturation term $S_{Ei}$ as a constant.
The root SCC $\mathcal{R}_s=\{E'_q,E'_d,\delta,\omega\}\subseteq\{\bm{x}\}$. To decide which of these variables, alone or in combination, should be measured, we build the symbolic Jacobian
\begin{equation}\label{eq:gen_symbolic_jac}
\widetilde{\bm{J}}=
\begin{bmatrix}
1&0&\bar{1}&0&1&0&0\\
0&1&\bar{1}&0&0&0&0\\
0&0&0&1&0&0&0\\
\bar{1}&\bar{1}&\bar{1}&1&0&0&0\\
0&0&0&0&1&0&1\\
0&0&0&0&1&1&0\\
0&0&0&0&1&1&1
\end{bmatrix},
\end{equation}
which has 12 constant and 5 nonlinear entries. Evaluating $\eta$ for each candidate output in $\mathcal{R}_s$, yields
\begin{equation}\label{eq:gen_eta}
\begin{aligned}
\eta_{E'_q}&=0.3626, & \eta_{E'_d}&=0.2653,\\
\eta_{\delta}&=0.1838, & \eta_{\omega}&=0.0021,
\end{aligned}
\end{equation}
and $\eta_{E_{fd}}=\eta_{R_f}=\eta_{V_R}=0$. Of the individual root-SCC variables, $E'_q$ scores highest. The gap between $E'_q$ and $E'_d$ comes from their dependency structures. The entry $\partial\dot E'_q/\partial E_{fd}=1/T'_{do}$ is a nonzero constant, while $\partial\dot E'_d/\partial E_{fd}=0$. Consequently, reconstructing from $E'_q$ picks up extra constant contributions from the excitation dynamics, which lifts its coefficient above that of $E'_d$ for this model. Multivariate measurements are handled by distributing the derivative orders among the outputs. 
For \(\bm{y}=[E'_q,E'_d,\delta]^\top\), the derivative orders are chosen as \([3,0,1]\), indicating that \(E'_q\) is differentiated three times, \(E'_d\) is not differentiated, and \(\delta\) is differentiated once. These yield seven coordinates for constructing the observability matrix with the rows of $E'_q$, $\dot{E'_q}$, $\ddot{E'_q}$, $\dddot{E'_q}$, $E'_d$, $\delta$ and $\dot{\delta}$. The resulting symbolic observability coefficient is \(\eta_{(E'_q,E'_d,\delta)}=1\), and the analytical observability determinant is
%For $\bm y=[E'_q\;E'_d\;\delta]^\top$ with derivative orders  $\eta_{(E'_q, E'_d, \delta)}=1$, and the analytical observability determinant
\begin{equation}\label{eq:gen_det}
\det\mathcal O_{(E'_q, E'_d, \delta)}
=\frac{K_A}{T_A T_E^2(T'_{do})^3},
\end{equation}
a nonzero constant for any physical parameter values. Swapping $\delta$ for $\omega$ under the same derivative allocation, its determinant in compact form is
\begin{equation}\label{eq:gen_det2}
\det\mathcal O_{(E'_q,E'_d,\omega)}
=-\frac{K_A}{T_AT_E^2(T'_{do})^3}g(\bm{x},\bm{u}).
\end{equation}
Unlike the previous combination, this determinant may vanish at certain operating points, reducing the symbolic observability coefficient to \(\eta=0.8776\). The combination of outputs matters, and so do the individual scores. However, in power systems, most state variables are not directly accessible. PMUs return terminal voltage and current synchrophasors rather than $\bm{x}$. The practical question is which physically available measurement maintains favorable symbolic coefficients while satisfying the root-SCC requirement. The next section works through that question on the WSCC nine-bus system.

% Replacing \(\delta\) with \(\omega\) under the same derivative allocation yields a state-dependent determinant, expressed as

% \[
% \det\mathcal O_{(E'_q,E'_d,\omega)}
% =-\frac{K_A}{T_AT_E^2(T'_{do})^3}g(\bm{x},\bm{u}),
% \]

% where \(g(\bm{x},\bm{u})=\partial f_\omega/\partial\delta\) is a nonconstant function of the system states and inputs. Unlike the previous combination, this determinant may vanish at certain operating points, reducing the symbolic observability coefficient to \(\eta=0.8776\).

\section{Numerical Validation}
\label{sec:results}
We evaluate the proposed methodology using the WSCC nine-bus system comprising three synchronous generators, as described in Example 7.1 of \cite{Sauer1997}. Each generator is represented by the seven-state nonlinear model in \eqref{eq:dif}, including the two-axis machine dynamics and IEEE Type-1 excitation system. The resulting centralized dynamic model consists of 21 states,

\begin{equation}
\bm{x} = \left[\bm{x}_1^\top \; \bm{x}_2^\top \; \bm{x}_3^\top \right]^\top \in \mathbb{R}^{21},
\end{equation}
where $\bm{x}_i^\top = \left[E'_{qi} \; E'_{di} \; \delta_i \; \omega_i \; E_{fdi} \; R_{fi} \; V_{Ri} \right]$. The network algebraic equations enter the generator dynamics through Kron reduction. The digraph built from the symbolic Jacobian of the centralized model has four SCCs, and a single root SCC $\mathcal{R}_s$ with 12 state variables \cite{Kandivalasa2026}:
\begin{equation}
\mathcal{R}_s =
\bigcup_{i=1}^{3}
\left\{
    E'_{qi}, E'_{di}, \delta_i, \omega_i
\right\}.
\label{eq:root_scc}
\end{equation}

To choose among these variables, we express the measured outputs as functions of the states. PMUs are considered at generator buses 1, 2, and 3. For a bus $i$,
\begin{equation}
\bm{y}_i = \left[V_{Di} \; V_{Qi} \right]^\top.
\label{eq:pmu_outputs}
\end{equation}
Combining the generator stator equations with the Kron-reduced network equations, we can express the outputs as $\bm{y} = \bm{h}(\bm{x})$. Every PMU output therefore depends on the root-SCC states of all three machines, that is,
\begin{equation}
\begin{aligned}
\bm{y}_1 &=  \left[V_{D1} \; V_{Q1} \right]^\top = \bm{h}_1(E'_{q1},E'_{d1},\delta_1,\ldots,
               E'_{q3},E'_{d3},\delta_3),\\
\bm{y}_2 &=   \left[V_{D2} \; V_{Q2} \right]^\top = \bm{h}_2(E'_{q1},E'_{d1},\delta_1,\ldots,
               E'_{q3},E'_{d3},\delta_3),\\
\bm{y}_3 &=  \left[V_{D3} \; V_{Q3} \right]^\top =  \bm{h}_3(E'_{q1},E'_{d1},\delta_1,\ldots,
               E'_{q3},E'_{d3},\delta_3).
\end{aligned}
\label{eq:pmu_functions}
\end{equation}

From these functions, we build symbolic observability matrices for individual voltage channels and for combinations across one, two, and three PMUs. Table~\ref{tab:pmu_eta} lists the resulting coefficients together with $\kappa$, the highest Lie-derivative order needed for the embedding to reach dimension 21. The proposed approach is computationally efficient because it replaces explicit nonlinear expressions with a symbolic representation of their dependencies. Rather than repeatedly computing and simplifying higher-order Lie derivatives, the method constructs successive rows of the symbolic observability matrix using predefined addition and multiplication rules over a limited set of symbols as in \eqref{eq:symbolicrules}. This avoids the rapid growth of analytical expressions and reduces the computational burden. 
The full analysis of the 21-state system takes under five seconds to run in MATLAB. For the same system, the Lie derivative-based approach takes 2 hours \cite{Kandivalasa2026}.
\begin{table}[t]
\centering
\caption{Symbolic Observability Coefficients for PMU Configurations}
\label{tab:pmu_eta}
\setlength{\tabcolsep}{5pt}
\begin{tabular}{lccc}
\toprule
\textbf{PMU Buses} & \textbf{Outputs} & $\kappa$ & $\eta$ \\
\midrule
1       & $\bm{y}_1$ & 10 & 0.0952 \\
2       & $\bm{y}_2$ & 10 & 0.0952 \\
3       & $\bm{y}_3$ & 10 & 0.0952 \\
1,2     & $\bm{y}_1,\bm{y}_2$       & 5  & 0.1905 \\
1,3     & $\bm{y}_1,\bm{y}_3$       & 5  & 0.1905 \\
2,3     & $\bm{y}_2,\bm{y}_3$       & 5  & 0.1905 \\
1,2,3   & $\bm{y}_1,\bm{y}_2,\bm{y}_3$   & 3  & 0.2857 \\
\bottomrule
\end{tabular}
\end{table}

Table~\ref{tab:pmu_eta} shows that $\eta$ increases with the number of measured PMUs. This is because of how the rows of $\widetilde{\mathcal{O}}$ are built. Each measured channel provides a row directly, without applying the multiplication and addition rules in~\eqref{eq:symbolicrules}, so that row keeps its simple symbols. Only the remaining rows must be generated through derivatives, and every derivative step combines symbols through these rules, which can only raise their algebraic complexity. With one PMU, two rows are direct, and nineteen must be generated through derivatives up to order $\kappa=10$. With three PMUs, six rows are direct, and the derivatives stop at $\kappa=3$. Fewer derivative steps leave more simple symbols in the determinant, which gives a larger $\eta$. Within each group, $\eta$ is the same regardless of location, because every PMU depends on the states of all three machines through the same symbolic pattern. The symbolic analysis therefore credits measurement channels but cannot tell the buses apart, or say which bus, or which pair of buses, reconstructs the state best.

The Lie-derivative analysis shows the same effect numerically. Deeper differentiation produces worse conditioning. At the nominal operating point, the condition numbers ($\sigma_{\min}/\sigma_{\max}$) of the Lie-derivative observability matrices range from $1.71\times10^{-14}$ to $1.82\times10^{-13}$ for single PMUs ($\kappa=10$), from $1.41\times10^{-8}$ to $3.28\times10^{-7}$ for pairs ($\kappa=5$), and reach $9.04\times10^{-6}$ for all three ($\kappa=3$). Conditioning improves with each added PMU, yet every matrix remains badly conditioned. The exact numbers depend on the model and on scaling, but the lesson holds: higher-order derivatives bring severe ill-conditioning even when the matrix has full rank. Measuring more channels reduces the number of derivatives needed, which is why both $\eta$ and the conditioning improve together.

The symbolic coefficients describe the algebraic structure of the system and carry no information about parameter values or operating conditions. Two configurations with the same coefficients can therefore estimate the states with different accuracy. We check this with an unscented Kalman filter (UKF). The UKF is run for seven measurement configurations: $\{1\}$, $\{2\}$, $\{3\}$, $\{1,2\}$, $\{1,3\}$, $\{2,3\}$, and $\{1,2,3\}$. Each PMU measures the direct- and quadrature-axis terminal voltages, $V_D$ and $V_Q$. Operating conditions and noise assumptions are identical across all configurations.

\begin{figure}[!t]
\centering
\includegraphics[width=\columnwidth]{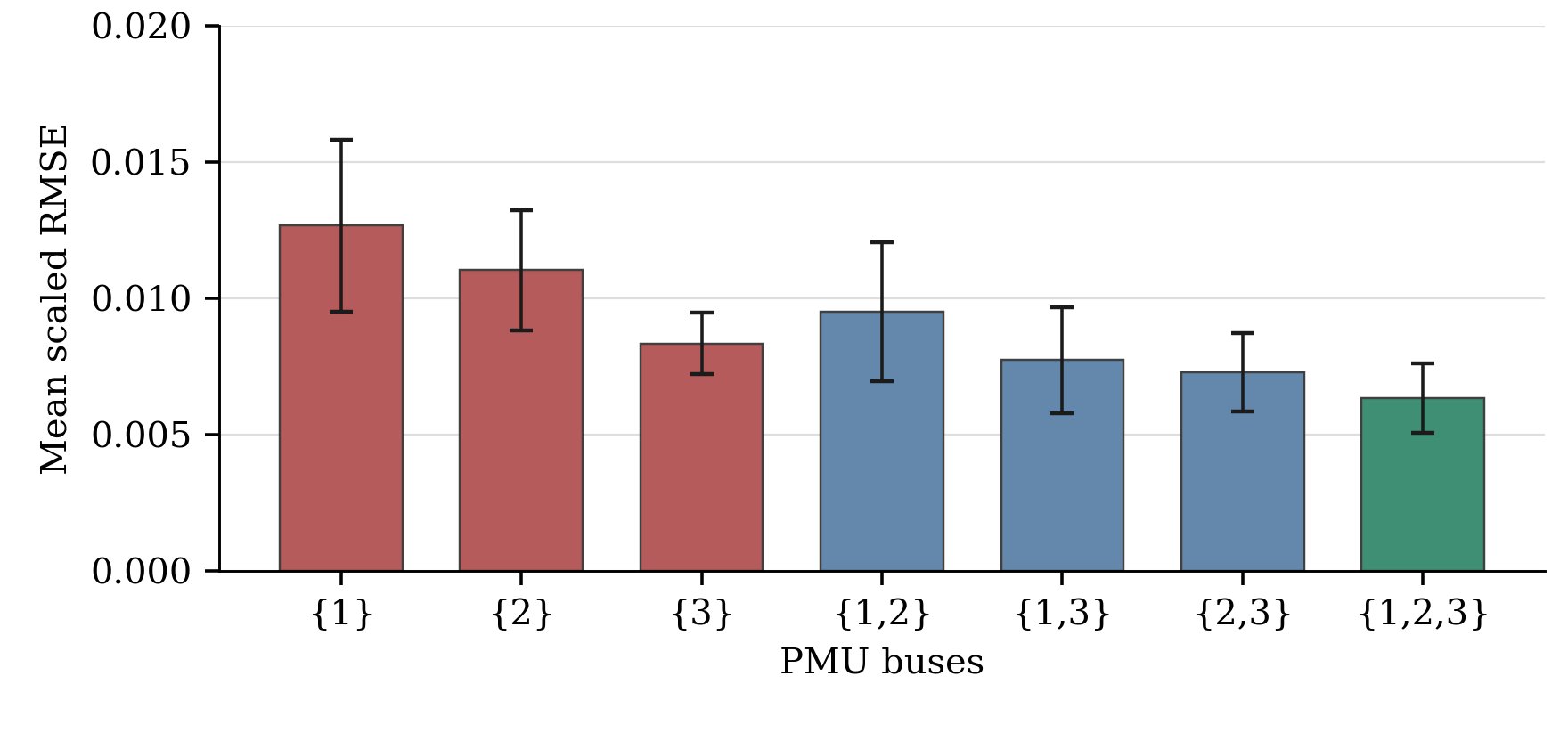}
\caption{Mean scaled RMSE for different PMU configurations in the WSCC nine-bus system. Error bars indicate one standard deviation across 12 trials.}
\label{fig:pmu_accuracy}
\end{figure}

The estimator is evaluated with a scaled root mean square error (RMSE). Because the 21 states have different units and magnitudes, we divide the RMSE of each state by the average magnitude of its true trajectory so we can average across states. For state $x_j$ over $N$ time steps,
\begin{equation}
\epsilon_j = \frac{\sqrt{\frac{1}{N}\sum_{k=1}^{N}\left(\hat{x}_j(k)-x_j(k)\right)^2}}{\frac{1}{N}\sum_{k=1}^{N}\left|x_j(k)\right|},
\label{eq:scaled_rmse}
\end{equation}
where $\hat{x}_j$ is the UKF estimate and $x_j$ the true state. The mean scaled RMSE is the average of $\epsilon_j$ over all 21 states. We run each configuration for 12 independent trials, and Fig.~\ref{fig:pmu_accuracy} reports the mean and standard deviation across them.

Fig.~\ref{fig:pmu_accuracy} shows that the estimation error depends on where the PMUs are placed, not only on how many. All three single-PMU configurations pass the same graph-based test and have the same $\eta$, yet their errors differ: Bus 3 gives the lowest mean scaled RMSE (0.00834), followed by Bus 2 (0.01103) and Bus 1 (0.01266). Among pairs, the two containing Bus 3, $\{1,3\}$ and $\{2,3\}$, perform better than $\{1,2\}$, whose mean error (0.0095) is higher than that of Bus 3 alone. Measuring all three buses gives the lowest error (0.0063). Adding PMUs generally reduces the error, but a well-placed single PMU can match or outperform a poorly placed pair, a distinction the symbolic coefficient cannot capture.

% The symbolic coefficient preserves the speed of the graph-based approach but does not reflect parameter values or operating conditions, as the UKF results show. Incorporating parameter-dependent measurement sensitivities into the graph, without losing its scalability, is the next step toward a measurement placement method for large-scale power systems.

\section{Conclusion}
\label{sec:conclusion}

This paper extended graph-based observability analysis for dynamic state estimation using symbolic observability coefficients to identify measurement selections that give stronger observability within a root SCC. For a seven-state synchronous machine, the coefficient varied with the choice of measured outputs. For the 21-state WSCC nine-bus system, it increased from 0.0952 with one PMU to 0.1905 and 0.2857 with two and three PMUs. The analysis took less than five seconds, compared with about two hours for the Lie-derivative approach. UKF results showed, however, that configurations with the same coefficient can differ in accuracy: among single PMUs, Bus 3 gave the lowest estimation error. 

The symbolic observability coefficients could not distinguish PMU locations because Kron reduction makes every output depend on all machines through the same symbolic pattern. Future work will retain the network equations instead of applying Kron reduction, so that the graph reflects the network topology and the electrical distance between each PMU and each machine. We will also incorporate parameter values and operating conditions to rank measurement configurations while keeping computational cost low.
\bibliography{lib}

@INPROCEEDINGS{Kandivalasa2026,
  author    = {Kandivalasa, Akhila and Netto, Marcos},
  booktitle = {2026 IEEE Power \& Energy Society General Meeting (PESGM)},
  title     = {Graph approach for observability analysis in power system dynamic state estimation},
  year      = {2026},
  pages     = {1--5},
  doi       = {10.1109/PESGM58988.2026.11693506}
}

@ARTICLE{Zhao2019,
  author  = {Zhao, Junbo and others},
  journal = {IEEE Trans. Power Syst.},
  title   = {Power System Dynamic State Estimation: Motivations, Definitions, Methodologies, and Future Work},
  year    = {2019},
  volume  = {34},
  number  = {4},
  pages   = {3188--3198}
}

@ARTICLE{Rouhani2017, author={Rouhani, Alireza and Abur, Ali}, journal={IEEE Trans. Power Syst.}, title={Observability Analysis for Dynamic State Estimation of Synchronous Machines}, year={2017}, volume={32}, number={4}, pages={3168-3175},}

@ARTICLE{Qi2015,
  author  = {Qi, Junjian and Sun, Kai and Kang, Wei},
  journal = {IEEE Trans. Power Syst.},
  title   = {Optimal {PMU} Placement for Power System Dynamic State Estimation by Using Empirical Observability {G}ramian},
  year    = {2015},
  volume  = {30},
  number  = {4},
  pages   = {2041--2054}
}

@book{Sauer1997, title={Power System Dynamics and Stability}, author={Sauer, Peter W. and Pai, M. A.}, year={1997}, publisher={Prentice Hall},}

@ARTICLE{KALMAN1960491,
  author  = {Kalman, R. E.},
  journal = {IFAC Proc. Vol.},
  title   = {On the general theory of control systems},
  year    = {1960},
  volume  = {1},
  number  = {1},
  pages   = {491--502},
  doi     = {10.1016/S1474-6670(17)70094-8}
}

@ARTICLE{Hermann_Krener_1977, author={Hermann, R. and Krener, A.}, journal={IEEE Trans. Autom. Control},  title={Nonlinear controllability and observability},  year={1977}, volume={22}, number={5}, pages={728-740}, doi={10.1109/TAC.1977.1101601}}

@article{Liu2013, author = {Yang-Yu Liu  and Jean-Jacques Slotine  and Albert-László Barabási}, title = {Observability of complex systems}, journal = {Proc. Natl. Acad. Sci.}, volume = {110}, number = {7}, pages = {2460-2465}, year = {2013},}

@ARTICLE{letellier2005relation,
  author  = {Letellier, Christophe and Aguirre, Luis A. and Maquet, Jean},
  journal = {Phys. Rev. E},
  title   = {Relation between observability and differential embeddings for nonlinear dynamics},
  year    = {2005},
  volume  = {71},
  number  = {6},
  pages   = {066213},
  doi     = {10.1103/PhysRevE.71.066213}
}

@ARTICLE{letellier2009symbolic,
  author  = {Letellier, Christophe and Aguirre, Luis A.},
  journal = {Phys. Rev. E},
  title   = {Symbolic observability coefficients for univariate and multivariate analysis},
  year    = {2009},
  volume  = {79},
  number  = {6},
  pages   = {066210},
  doi     = {10.1103/PhysRevE.79.066210}
}

@ARTICLE{letellier2018symbolic,
  author  = {Letellier, Christophe and Sendi{\~n}a-Nadal, Irene and Bianco-Martinez, Ezequiel and Baptista, Murilo S.},
  journal = {Sci. Rep.},
  title   = {A symbolic network-based nonlinear theory for dynamical systems observability},
  year    = {2018},
  volume  = {8},
  number  = {1},
  pages   = {3785},
  doi     = {10.1038/s41598-018-21967-w}
}

@book{Zhao2026,
author = {Junbo Zhao and Marcos Netto and Lamine Mili},
title = {Robust Dynamic State Estimation of Power Systems},
publisher = {Elsevier},
year = {2026},}
\bibliographystyle{ieeetr}

\end{document}